\documentclass[aps,prl,amsmath,amssymb,twocolumn,preprintnumbers,superscriptaddress,showpacs,lengthcheck]{revtex4-1}

\newcommand{\bscco}{$\textrm{Bi}_2\textrm{Sr}_2\textrm{CaCu}_2\textrm{O}_{8+\delta}$}
\newcommand{\cecoin}{$\textrm{Ce}\textrm{Co}\textrm{In}_5$}
\newcommand{\urusi}{$\textrm{U}\textrm{Ru}_2\textrm{Si}_2$}

\usepackage{graphicx}
\usepackage{dcolumn}
\usepackage{bm}
\usepackage{pdfpages}
\pdfoutput=1

\begin{document}



\title{Detection of electronic nematicity using scanning tunneling microscopy}


\author{Eduardo H. da Silva Neto}
\affiliation{Joseph Henry Laboratories and Department of Physics, Princeton University, Princeton, NJ 08544, USA}

\author{Pegor Aynajian}
\affiliation{Joseph Henry Laboratories and Department of Physics, Princeton University, Princeton, NJ 08544, USA}

\author{Ryan E. Baumbach}
\affiliation{Los Alamos National Laboratory, Los Alamos, New Mexico 87545, USA}

\author{Eric D. Bauer}
\affiliation{Los Alamos National Laboratory, Los Alamos, New Mexico 87545, USA}

\author{John Mydosh}
\affiliation{Kamerlingh Onnes Laboratory, Leiden University, 2300 RA Leiden, The Netherlands}

\author{Shimpei Ono}
\affiliation{Central Research Institute of Electric Power Industry, Komae, Tokyo, Japan}

\author{Ali Yazdani}
\email{yazdani@princeton.edu}
\affiliation{Joseph Henry Laboratories and Department of Physics, Princeton University, Princeton, NJ 08544, USA}



\begin{abstract}
Electronic nematic phases have been proposed to occur in various correlated electron systems and were recently claimed to have been detected in scanning tunneling microscopy (STM) conductance maps of the pseudogap states of the cuprate high-temperature superconductor \bscco~(Bi-2212). We investigate the influence of anisotropic STM tip structures on such measurements and establish, with a model calculation, the presence of a tunneling interference effect within an STM junction that induces energy-dependent symmetry-breaking features in the conductance maps. We experimentally confirm this phenomenon on different correlated electron systems, including measurements in the pseudogap state of Bi-2212, showing that the apparent nematic behavior of the imaged crystal lattice is likely not due to nematic order but is related to how a realistic STM tip probes the band structure of a material. We further establish that this interference effect can be used as a sensitive probe of changes in the momentum structure of the sample's quasiparticles as a function of energy.

\end{abstract}


\maketitle


The concept of broken symmetry is essential to condensed matter physics. Identification of the fundamental symmetries of a solid-state system leads to the understanding of the low-energy excitations which govern its properties. For example, the unraveling of the three-decade old mystery of unconventional superconductivity hinges on determining the symmetries of the correlated electronic state from which cooper pairs are formed. Recently, electronic nematic phases, where electronic states undergo a spontaneous four-fold ($C_{4}$) to two-fold ($C_{2}$) symmetry breaking, have gained much interest as a possible candidate for various \emph{hidden order} states in several correlated electron systems such as cuprates, iron-based superconductors, and heavy fermion materials \cite{kivelson_how_2003, fradkin_nematic_2010, daou_broken_2010, Chu_2010, okazaki_torque, Chu_2012}. However, such states are difficult to detect using non-local probes because of possible twin-domain structures in macroscopic samples. Recently STM has been proposed as the method of choice for the detection of four-fold electronic symmetry breaking \cite{Chuang_nematic_2010, lawler_intra-unit-cell_2010, Mesaros_2011, Jenny_structural_2012, hamidian_NJP_2012, Song_2012, kohsaka_2012, Jenny_FeSe_2012}. Lawler \emph{et al.} \cite{lawler_intra-unit-cell_2010}, and Messaros \emph{et al.} \cite{Mesaros_2011} in a subsequent study, interpreted the rotational symmetry breaking in the STM data as evidence for an electronic nematic state inside the pseudogap phase of Bi-2212. 

The na\"{\i}ve expectation has been that the influence of the STM tip's geometric structure is limited to inducing an easy-to-detect anisotropy in STM topographs or to influence energy-resolved STM differential conductance ($dI/dV$) maps in an energy independent manner. Here we show, through model calculations and experimental measurements on three correlated electron materials (\cecoin, Bi-2212, and \urusi), that a tunneling interference effect within an STM junction composed of a realistic tip (with some spatial anisotropy) can result in an artificial energy-dependent symmetry breaking of the STM conductance maps. This phenomenon can occur even when the STM topograph taken with the same tip appears to be symmetric. We demonstrate that previously reported two-fold symmetric conductance maps in high-$T_c$ cuprates \cite{lawler_intra-unit-cell_2010, Mesaros_2011} are not evidence for rotational symmetry breaking ($C_{4}$ to $C_{2}$) originating from a nematic phase in these materials, but are rather due to the interference effect we have uncovered here. In this system, systematic measurements with different tips on the same area of the sample, reported here, are also used to clearly demonstrate the lack of nematic order, without relying on any pseudogap-specific assumptions about the tunneling process.  We further show that the interference effect within the STM junction can nevertheless be used as a sensitive tool to detect changes in the quasiparticle band structure as a function of energy.

\begin{figure}
\includegraphics[width=84mm]{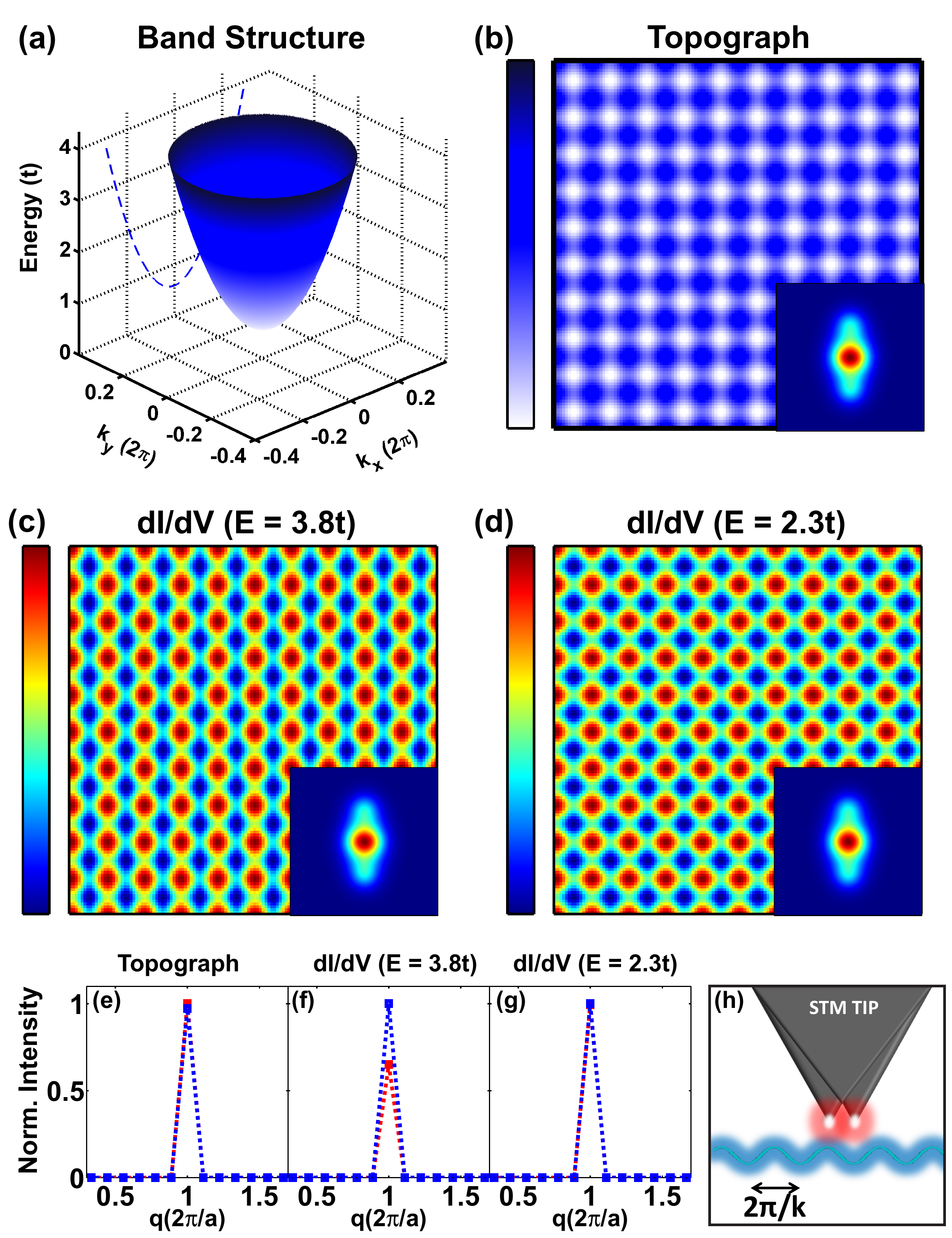}
\caption{\label{fig:1} (Color online) (a) Parabolic band structure ($E = E_0 + t k^2 $, with $t=(50/\pi^2)$ and $E_0 =0.59 t$) used to generate (b), (c) and (d). (b) Simulated topography at $eV=E=3.95t$ imaged by the tip in the lower inset. The inset of the tip represents the simulated $|\psi_{t}|^{2}$ and is plotted on the same spatial scale as the lattice. (c) Simulated differential conductance at $E=3.8t$ showing rotational symmetry breaking with the same tip as in (b). (d) Simulated differential conductance at $E=2.3t$ showing no rotational symmetry breaking with the same tip as in (b). (e-g) represent the intensity (normalized to the maximum) of the two orthogonal Bragg peaks generated by the DFTs of (b), (c), and (d), respectively. (h) One-dimensional schematic representation of the interference between the wavefunction of a double tip with a quasiparticle state of momentum $\vec{k}$.}
\end{figure}

We start our discussion by considering how STM probes the electronic structure of a sample's surface.  Following Tersoff and Hamann \cite{tersoff_hamann_1985} the sample wave function can be written as
\begin{eqnarray} \label{eq:1}
\psi_{s,\vec{k}} (\vec{r}) &=& \sum_{\vec{G}} a_{\vec{G}} exp[i \vec{\kappa}_{G} \cdot \vec{r} -(\kappa^2 + |\vec{\kappa}_{G}|^2)^{\frac{1}{2}} z]
\end{eqnarray}
where $\vec{\kappa}_{G} = \vec{k} + \vec{G}$ defines the Wannier states, while $\kappa = \sqrt{2m\phi}/\hbar$ is related to the work function $\phi$ for electron decay into the vacuum, and the summation is over the reciprocal lattice vectors $\vec{G}$ \cite{SM}. Most discussions of STM data assume a metalic tip (energy-independent density of states), and approximate the STM differential conductance $dI/dV$ (at small bias) as the spatial convolution ($*$) of the tip ($\rho_t$) and sample ($\rho_s$) densities of states (DOS) \cite{tersoff_hamann_1985}:
\begin{eqnarray}
\label{eq:2}
\frac{dI}{dV}(eV,\vec{r}) \propto \rho_{t} *\rho_{s}(eV)
\end{eqnarray}
with $\rho_{s}(eV) = \sum_{s}|\psi_{s}|^{2} \delta (E_{s} - E_f - eV)$. Then under the usual assumption of an isotropic tip, the conductance maps are simply proportional to the local DOS of the sample. Within this model of STM measurements the topograph is constructed from integrating such maps between the Fermi level up to the tip-sample bias.

Before considering an anisotropic tip, we emphasize that Eq.\,(\ref{eq:2}) above is an approximation of Bardeen's formula for tunneling \cite{bardeen_1961,tersoff_hamann_1985, SM}
\begin{eqnarray}
\label{eq:3}
I= \frac{e}{h} \sum_{s,t} f(E_s)[1 - f(E_t + eV)]|M_{s t}|^2 \delta(\Delta E)
\end{eqnarray}
where $f(E)$ is the Fermi-Dirac distribution, and $\Delta E = E_s - E_t$. Here $M_{st}$ is the matrix element between the tip and the sample with the following spatial structure \cite{bardeen_1961}: 
\begin{eqnarray}
\label{eq:4}
M_{s t} = \frac{\hbar^2}{2 m} \int d\vec{\bf{S}} \cdot (\psi_s^* \vec{\nabla}\psi_t - \psi_t \vec{\nabla}\psi_s^*)
\end{eqnarray}

We now demonstrate the effect of the tip geometry on STM measurements, by simulating the conductance maps using Bardeen's matrix element (Eq.\,(\ref{eq:4}) above) with a two-fold symmetric tip structure, characterized by orthogonal lengths $\delta x = 0.2$, $\delta y = 0.9$ (lattice constant set to unity) \cite{SM}. We assume that the sample has a $C_{4}$ symmetric electronic structure with a generic parabolic band structure Fig.\,\ref{fig:1}(a), which is isotropic in the $k_x$\,-\,$k_y$ plane. Calculations of the STM topographic image (at $eV=3.95t$) and conductance maps at two different energies of this four-fold symmetric sample with elongated tip wave function are shown in Fig.\,\ref{fig:1}(b-d). While the STM conductance maps can show apparent asymmetry in the $x$\,-\,$y$ plane Fig.\,\ref{fig:1}(c), the topograph appears to be remarkably four-fold symmetric Fig.\,\ref{fig:1}(b). Clearly, the summation of $dI/dV$ maps over an appropriate range of energies can lead to a four-fold symmetric topograph. This finding demonstrates that a four-fold symmetric topograph cannot be used as an accurate method to characterize the STM tip geometry, as it is often assumed (for example see Ref.\,\cite{lawler_intra-unit-cell_2010}).  

An anisotropic tip would naturally induce an apparent breaking of four-fold symmetry in measurements of the electronic structure of a four-fold symmetric sample, as the conductance maps demonstrate. However, the energy dependence of the $x$-$y$ asymmetry (which can change sign, see below), or its absence for some energies (Fig.\,\ref{fig:1}(d)), points to a previously overlooked interference effect of STM measurements. Examining Eq.\,(\ref{eq:4}) we realize that the periodic corrugation along the $x$ and $y$ directions in the conductance maps are determined by the interference (see Fig.\,\ref{fig:1}(h) for schematic) between the sample's quasiparticle states $\psi_{s,\vec{k}}$ with momentum $\vec{k}$ together with those of the tip (characterized in our two-fold symmetric tip by $\delta x$ and $\delta y$). Previous studies of the influence of asymmetric tips \cite{PhysRevB.35.7790, snyder_1990, PhysRevB.57.13118} have overlooked the interference within the STM junction by using the approximation in Eq.\,(\ref{eq:2}) which ignores the phase information ($e^{i \vec{k} \cdot \vec{r}}$, see Eq.\,(\ref{eq:1})) of the sample wave functions that are relevant in the evaluation of Eq.\,(\ref{eq:4}). Additionally, studies of quantum interference effects in tunneling junctions \cite{PhysRevB.68.115425, Jurczyszyn2003185, Nieminen2004L47,khotkevych:53} have not considered the effects of geometrically asymmetric tips on the measurement of long-range periodic structures by the STM. In contrast, our model calculations clearly show that the electronic structure of a four-fold symmetric square lattice probed by a real STM tip can be two-fold symmetric depending on the momentum $\vec{k}$ (consequently energy) of the quasiparticles probed.

\begin{figure}
\includegraphics[width=87mm]{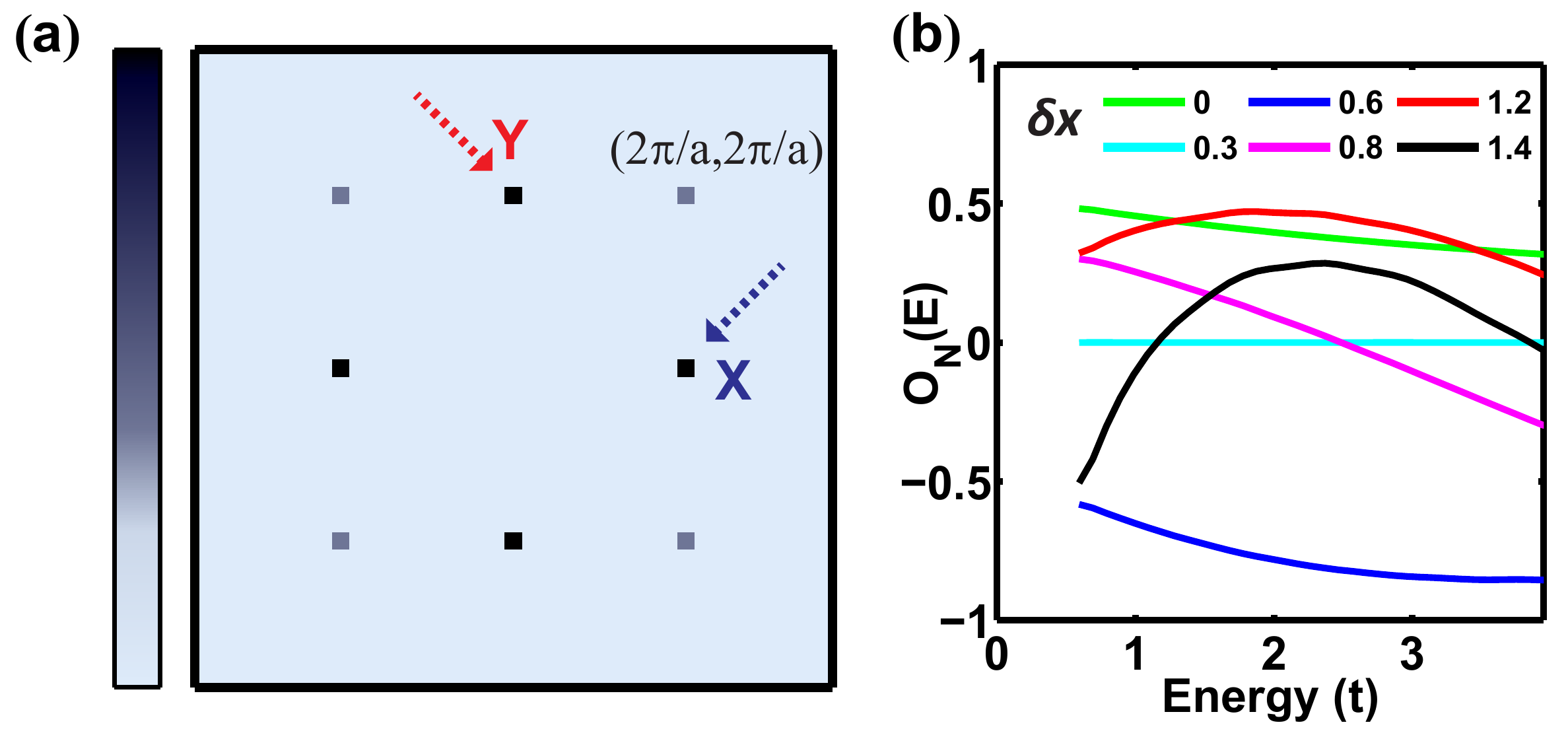}
\caption{\label{fig:2} (Color online) (a) DFT of $dI/dV (E=0.79t)$ generated using Eq.\,(\ref{eq:2}) showing strong peaks due to the long-range periodic lattice strucutre. (b) Energy dependence of the lattice asymmetry parameter calculated from Eq.\,(\ref{eq:7}) for different tip configurations ($\delta y = 0.3$, and $\delta x$ as indicated in the figure). Notice that for a four-fold symmetric tip ($\delta x = 0.3$ curve in (b)) $O_N(E) = 0$ for all energies.}
\end{figure}

For a more detailed analysis of the energy-dependence of this asymmetry in STM conductance maps, and to make a  connection to experimental measurements, we quantify the calculated STM conductance maps with the two-dimensional asymmetry parameter that is commonly used in the context of nematic ordering \cite{fradkin_nematic_2010}: 
\begin{eqnarray}
\label{eq:7}
O_{N}(E) = \frac{X(E)-Y(E)}{X(E)+Y(E)}
\end{eqnarray}
where $X(E)$ and $Y(E)$ are the energy dependent amplitudes of the two Bragg peaks along the orthogonal directions obtained from discrete Fourier transformation (DFT), as indicated in Fig.\,\ref{fig:2}(a). A map with $O_{N} = 0$ corresponds to a four-fold symmetric image, whereas $O_{N} = 1$ indicates an image with zero corrugation along either the $x$ or $y$ direction (as expected for example for one-dimensional stripes). Figure \ref{fig:2}(b) shows $O_{N}(E)$ for different tip configurations. Despite the simplicity of the model band structure, $O_{N}(E)$ shows a significant energy dependence over the entire bandwidth and even a sign change. This illustrates the sensitivity of this tip-induced interference effect to the band structure. Although the magnitude of $O_{N}(E)$ can only be understood by a detailed knowledge of $\psi_{t}$, its energy dependence acts as a detector of changes in the momentum structure of the quasiparticle states.

\begin{figure}
\includegraphics[width=87mm]{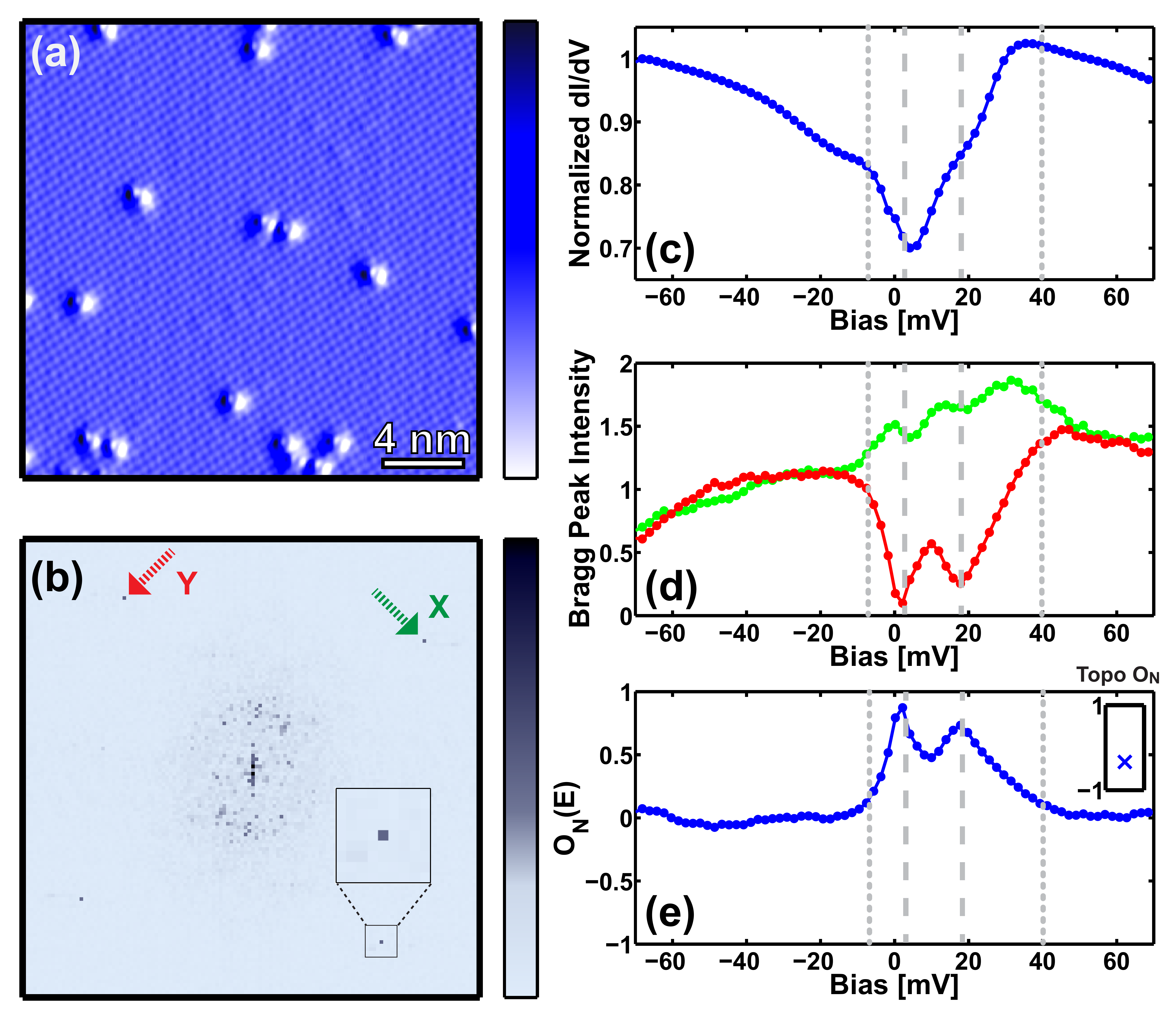}
\caption{\label{fig:3} (Color online) (a) Topograph of \cecoin~(setpoint condition at $-200$\,mV and $1.6$\,nA) showing a square lattice. For enhanced contrast the derivative of the data is shown. (b) DFT of conductance ($dI/dV$) map taken at $-52$\,meV over the same FOV as (a) showing strong Bragg peaks representing the square lattice. Inset shows an enlargement of the bottom right Bragg peak. (c) Tunneling spectrum averaged over the area in (a). (d) Energy dependence of the Bragg peak intensity obtained via the DFT operation. (e) Asymmetry parameter calculated via Eq.\,(\ref{eq:7}). Inset of (e) represents the asymmtery parameter of the topograph acquired simultaneously to the conductance maps.}
\end{figure}

Our model calculation suggests that materials with changes in their electronic band structure as a function of energy (such as a rapid change of band dispersion) are likely to be good candidates for exhibiting the interference effect associated with asymmetric tips. A good material candidate for such a study is the heavy fermion compound \cecoin, which crystalizes in the tetragonal crystal structure, ensuring the four-fold symmetry of its electronic states. Recent STM studies on \cecoin~have demonstrated that the electronic structure of this compound exhibits the development of a hybridization gap and associated heavy bands near the Fermi energy at low temperatures \cite{aynajian_visualizing_2012}. Figure\,\ref{fig:3} shows a topograph (a) of \cecoin, a DFT of a conductance map on the same area (b) together with the STM spectrum as a function of energy (c), which demonstrates the presence of a hybridization gap in this compound near the Fermi energy. Also shown in Fig.\,\ref{fig:3} is the intensity of the Bragg peaks in the conductance maps as a function of energy (d) and the asymmetry parameter $O_{N}(E)$ (e) introduced above.

Approaching the energy window near the Fermi level, where we expect strong changes in the band structure of \cecoin~due to hybridization of \emph{spd}- and \emph{f}-like electrons, we find a strong $C_4$-symmetry breaking of the STM conductance maps (Fig.\,\ref{fig:3}(d,e)). Both the Bragg peaks and $O_{N}(E)$ show an energy-dependent asymmetric behavior in tandem with the features of the STM spectrum. The apparent breaking of $C_{4}$ symmetry in this experiment is not associated with nematic order in \cecoin~but rather probes the strong momentum dependence of the band structure of this compound near the Fermi level. In fact, it is remarkable that $O_N(E)$ is sensitive to the most subtle features in the spectra as a function of energy (see dashed lines in Fig.\,\ref{fig:3}(c)), which are associated with changes in the electronic momentum structure as previously detected in the quasiparticle interference of this compound \cite{aynajian_visualizing_2012}.

\begin{figure*}
\includegraphics[width=160mm]{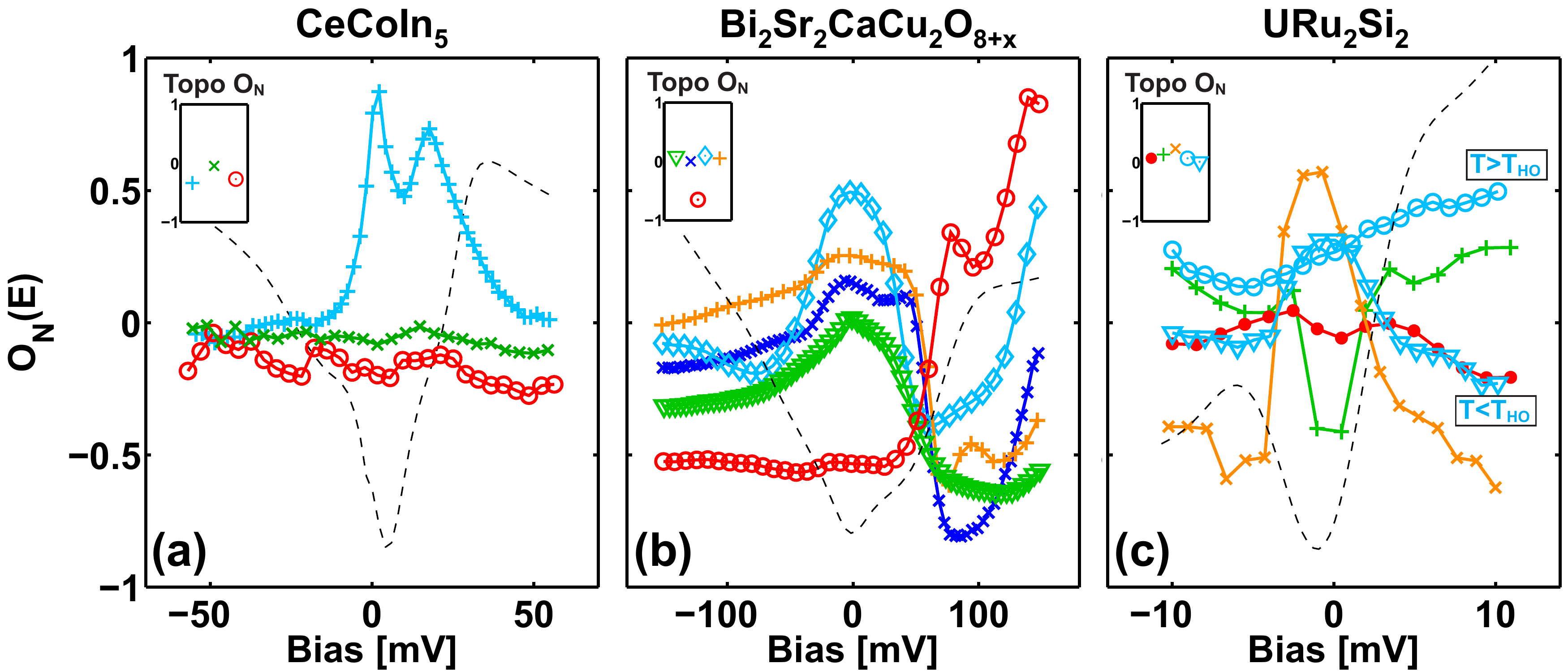}
\caption{\label{fig:4} (Color online) (a) $O_{N}(E)$ measured on \cecoin~with different tips at $20$\,K. (b) $O_{N}(E)$ measured on $\textrm{Bi}_2\textrm{Sr}_2\textrm{(Ca,Dy)Cu}_2\textrm{O}_{8+\delta}$ ($T_c = 35$\,K) at $30$\,K over the same FOV with different tip configurations. (c) Open symbols represent $O_{N}(E)$ measured on \urusi~above $T_{HO}$ ($20$\,K) and below ($15$\,K) over the same FOV with the same tip. Measurements on a second FOV with different tip configurations (closed symbols) were carried out below $T_{HO}$ ($15$\,K). For comparison purposes, the average tunneling spectra are displayed (dashed curves) for the respective materials (for \urusi~the average spectrum at $10$\,K is displayed) \cite{SM}. The insets represent the asymmtery parameter of the respective individual topographs acquired simultaneously to the conductance maps.}
\end{figure*}

Further evidence that the asymmetry between $X$ and $Y$ detected in the conductance maps of \cecoin~is associated with interference in the STM junction can be found by repeating the same experiment with slightly different tips (created by interacting with the surface) over the same field of view (FOV) or equivalent areas of the same cleaved sample. As expected from our model calculations (Fig.\,\ref{fig:2}(b)) different tips have different sensitivity to the momentum structure of the electronic structure of the sample, and depending on their geometry exhibit different degrees of $C_4$-symmetry breaking in the acquired conductance maps. As Fig.\,\ref{fig:4}(a) shows, the energy dependence of the asymmetry parameter in the conductance maps, $O_{N}(E)$, is a very strong function of the tip and not always correlated with the presence, or the degree of, Bragg peak asymmetry in the STM topographs of the same area.

We turn our attention next to the claims that STM measurements of underdoped Bi-2212 samples break $C_4$ symmetry and exhibit nematic order \cite{lawler_intra-unit-cell_2010, Mesaros_2011}. As shown in Fig.\,\ref{fig:4}(b) measurements on such a sample exhibit very similar characteristics to those of \cecoin, where changes in the spectrum associated with the pseudogap coincide with apparent asymmetry and a non-zero $O_{N}(E)$ in this energy window. Not only this correlation is very characteristic of the tip-induced symmetry breaking originating from interference effects within the STM junction, we also find that $O_{N}(E)$ displays a strong sensitivity to the tip structure when probing the exact same FOV with slightly different tips. Remarkably, tips that show very similar, nearly $x$-$y$ symmetric, topographs can show very different energy dependences for $O_{N}(E)$, and even exhibit opposite signs for the effect on the same exact area of the sample. Clearly, such behavior is more consistent with the tip-dependent interference in the STM junction, associated with changes in the momenta of electronic states within the pseudogap, rather than any nematic order. Consistent with this view, and with previous experiments \cite{lawler_intra-unit-cell_2010, Mesaros_2011}, no domain boundaries between regions of different nematic order parameter have ever been found despite the large areas used for STM studies.

Before we conclude, we present experiments on one more materials system, the results of which demonstrate that the interference within the STM junction and the associated asymmetry parameter can in fact be used to probe the onset of sudden changes in electronic band structures of materials. We have carried out temperature dependent experiments on the heavy fermion \urusi, which shows a sharp second order phase transition in the so-called "hidden order state" below $T_{HO}=17.5$\,K, the nature of which continues to be a mystery \cite{palstra_superconducting_1985, aynajian_visualizing_2010, schmidt_imaging_2010}. Experiments on this compound are also consistent with the asymmetry parameter picking up changes in the electronic states at low temperatures through the tip-dependent interference. However, contrasting measurements below and above the $T_{HO}$, over the same FOV, and with the same tip (open symbols in Fig.\,\ref{fig:4}(c)), shows that the signals in $O_{N}(E)$ change from a peak-like shape to a smooth curve, directly reflecting the change in the band structure as the hidden order phase transition is crossed. At temperatures just below the transition, when the changes in the electronic states are difficult to detect in the STM spectra, we find that $O_N(E)$ is extremely sensitive to the changes that occur in the electronic structure of this material below $T_{HO}$.

Overall, our systematic measurements on three different materials (with three different characteristic gap energy scales, $30$\,meV for \cecoin, $4$\,meV for \urusi, and $100$\,meV for Bi-2212), demonstrate the strong sensitivity of the asymmetry parameter $O_{N}(E)$ to different tip configurations, and, specifically, how it can change sign for measurements over the same FOV. From these results, we can only conclude that $O_{N}(E)$ is not a measure of the symmetry breaking of the electronic states of the sample, rather it is the result of the interference effect which is evident in the elementary model of tunneling from a realistic tip discussed in here. Although STM can in principle detect the onset of nematic order, we have demonstrated that symmetry analyses of conductance maps can be dominated by the energy-dependence of the band structure of the sample rather than nematic order. Perhaps the only experimentally reliable approach to detect rotational symmetry breaking order with the STM would be to image the presence of domain boundaries, such as the structural ordering \cite{Jenny_structural_2012} and electronic smectic ordering \cite{Mesaros_2011} in Bi-2212, and electronic nematic ordering in iron-based superconductors \cite{Jenny_FeSe_2012, Chuang_nematic_2010}. Alternatively rotation of the STM tip by an appropriate angle ($90^{\circ}$ in the case of $C_{4}$ symmetry) while maintaining the same location on the sample could be developed to discount the role of the tip geometry. Regardless, the interference within the STM junction with realistically anisotropic tips described here shows that such measurements are a sensitive probe of the changes in the quasiparticle states of the sample as a function of energy, even when such changes might not be apparent in STM spectra.

\begin{acknowledgments}
Work at Princeton University was primarily supported by a grant from the DOE Office of Basic Energy Sciences (DE-FG02-07ER46419). The instrumentation and infrastructure at the Princeton Nanoscale Microscopy Laboratory are also supported by grants from the NSF-DMR1104612 and NSF-MRSEC programmes through the Princeton Center for Complex Materials (DMR-0819860). Work at LANL was conducted under the auspices of the US DOE Office of Basic Energy Sciences, Division of Materials Science and Engineering.
\end{acknowledgments}

\bibliographystyle{apsrev4-1}
\bibliography{bib_lib}
\clearpage


\section*{Supplementary material}

# Detection of electronic nematicity using scanning tunneling microscopy
## *Supplemental Material*

Eduardo H. da Silva Neto,[1] Pegor Aynajian,[1] Ryan E. Baumbach,[2]
Eric D. Bauer,[2] John Mydosh,[3] Shimpei Ono,[4] and Ali Yazdani[1, *]

[1]*Joseph Henry Laboratories and Department of Physics, Princeton University, Princeton, NJ 08544, USA*
[2]*Los Alamos National Laboratory, Los Alamos, New Mexico 87545, USA*
[3]*Kamerlingh Onnes Laboratory, Leiden University, 2300 RA Leiden, The Netherlands*
[4]*Central Research Institute of Electric Power Industry, Komae, Tokyo, Japan*

## THE BARDEEN MATRIX ELEMENT

In Bardeen's formalism [1, 2] the tunneling current is determined by

$$I = \frac{2\pi e}{\hbar} \sum_{s,t} f(E_s)[1 - f(E_t + eV)]|M_{st}|^2 \delta(E_s - E_t) \tag{1}$$

where $s(t)$ indexes the sample (tip), $f(E)$ is the Fermi-Dirac distribution, and $M_{st}$ is a matrix element between the sample and tip states. The spatial structure of the STM image is therefore contained in $M_{st}$, which Bardeen [1] showed to be

$$M_{st} = \frac{\hbar^2}{2m} \int d\vec{S} \cdot (\psi_s^* \vec{\nabla} \psi_t - \psi_t \vec{\nabla} \psi_s^*) \tag{2}$$

where $\vec{S}$ is any closed surface in the vacuum separating the tip and the sample. We choose the surface of the sample to be the $z = 0$ plane. Then the tip is above at coordinates $(x_t, y_t, z_t)$, and $\vec{S}$ can be the $0 < z < z_t$ plane. Manipulation of Eq. (2) shows that $M_{st}$ resembles a probability current through the $z$ surface:

$$M_{st}(x_t, y_t, z_t) = \frac{\hbar^2}{2m} \int \left( \psi_s^*(x, y, z) \frac{\partial \psi_t(x - x_t, y - y_t, z - z_t)}{\partial z} - \psi_t(x - x_t, y - y_t, z - z_t) \frac{\partial \psi_s^*(x, y, z)}{\partial z} \right) dx \, dy \tag{3}$$

It can now be seen explicitly that the matrix element is dependent on the position of the tip. Equation (3) describes the spatial structure of the STM measurement and it can be written in a simplified way:

$$M_{st}(x_t, y_t, z_t) = \frac{\hbar^2}{2m} \left( \psi_s^* * \frac{\partial \psi_t}{\partial z} - \psi_t * \frac{\partial \psi_s^*}{\partial z} \right), \tag{4}$$

where the $*$ operation denotes convolution in the $x$ and $y$ coordinates, and $M_{st}(x_t, y_t, z_t)$ is a function of the tip coordinates.

## SAMPLE AND TIP WAVE FUNCTIONS

### Sample

According to Tersoff and Hamann [2] the sample wave function is written as

$$\psi_{s,\vec{k}}(\vec{r}) = \sum_{\vec{G}} a_{\vec{G}} exp[i\vec{\kappa}_G \cdot \vec{r} - (\kappa^2 + |\vec{\kappa}_G|^2)^{\frac{1}{2}} z] \tag{5}$$

$$\vec{\kappa}_G = \vec{k} + \vec{G} \tag{6}$$

$$\kappa = \frac{(2m\phi)^{\frac{1}{2}}}{\hbar} \tag{7}$$

The $z$-dependence depends on the work function $\phi$ and on $\vec{G}$. For large $\vec{G}$ the penetration of the corresponding Bloch waves into the vacuum decreases exponentially. Therefore we restrict the sum in Eq. (5) to the three highest order terms: $\vec{G} = 0$, $\vec{G} = \vec{G_x}$, and $\vec{G} = \vec{G_y}$, where without loss of generality we have taken the simplest case of a square lattice defined by reciprocal lattice vectors $\vec{G_x}$ and $\vec{G_y}$. These three highest order terms also represent the simplest wave function which can produce a square lattice in Eq. (5). Here we have taken the values $a_0 = 1$, and $a_{\vec{G_x}} = a_{\vec{G_y}} = 2e^{i\frac{\pi}{4}}$. Though these coefficients determine the spatial phase of the periodic lattice structure, different relative values have been tested and do affect our conclusions. The results of the present study can also be easily expanded to other Bravais lattices, or other long-range periodic structures, by appropriate choice of the reciprocal space vector basis.

### Tip

First we consider an ideal tip, where the characteristic scale of its wave function is much smaller than the period of $\psi_s$, then $\psi_t$ can be modeled as an infinitely sharp delta function, and Eq. (5), together with Eq. (4), yield $M_{st} \propto |\psi_s|^2$ and consequently $dI/dV \propto \rho_s$. This is the usual STM assumption that the differential conductance

is proportional to the local density of states of the sample. However, Tersoff and Hamann [2] showed that an infinetely sharp tip is not necessary to obtain atomic resolution with the STM. We extend their results to accomodate for anisotropic tips. Considering the typical separation between sample and tip ($\approx 5$ Å) we model the tip as an s-wave:

$$\psi_t = e^{i\alpha}e^{-\Gamma(x^2+y^2+(z-z_t)^2)^{\frac{1}{2}}}, \qquad (8)$$

where $\alpha$ is the phase of this wave function and $\Gamma$ determines its penetration into the vacuum [3]. We find that the simplest choice to model an anisotropic tip is to write it's wave function as a combination of four s-waves with their centers separated by $2 \times \delta x$, and $2 \times \delta y$:

$$\psi_t = e^{-\Gamma((x-\delta x)^2+y^2+(z-z_t)^2)^{\frac{1}{2}}}$$
$$+ e^{-\Gamma((x+\delta x)^2+y^2+(z-z_t)^2)^{\frac{1}{2}}} + (x \leftrightarrow y) \qquad (9)$$

We have chosen this particular functional form for $\psi_t$ because of its simplicity, though any other choice of anisotropic wave function in the $x$-$y$ plane suffices to produce the apparent rotational symmetry breaking described in the text.

## CALCULATION STRATEGY

The caculation is performed in the following steps. First, for each constant energy contour of the band structure in Fig. 1, the momentum states $\vec{k}$ are represented by a circle in the $k_x$-$k_y$ plane. We then slice that circle in $N$ points equally spaced by angle, which are represented by unique values of $k_x$ and $k_y$. For each pair we calculate $M_{st}$ according to Eq. (4). Finally we add all the $|M_{st}|^2$ (where the $s$ states in the summation of Eq. (1) are the several $\psi_{s,\vec{k}}$ states for a particular energy) to obtain the conductance map at the energy of interest. We repeat this procedure for each energy and construct the curves in Fig. 2(a) of the main text.

## ANGLE BETWEEN TIP ANISOTROPY AXIS AND THE LATTICE

In Eq. 9 we implicitely assume that the anisotropy axis of the tip wavefunction is aligned with one of the crystal axes. This is rarely the case in an actual experiment. However, for most values of relative angles between the tip anisotropy axis and the crytal lattice, the induced $x$-$y$ asymmetry in the conductance maps should still occur. We confirm this expectation by performing our calculation for different relative angles (see Fig. 1). Notice that for a value of $45°$ the tip geometry is unable to distinguish between the $x$ and $y$ axes of the sample, and therefore the value of $O_N(E)$ is identically zero for all energies.

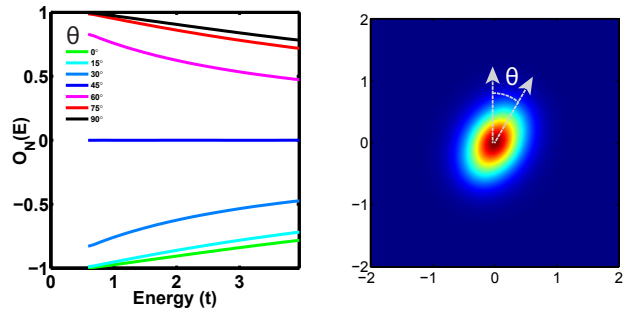

FIG. 1. $O_N(E)$ (left panel) for select angles between the crystal lattice (vertical arrow, right panel) and the tip anisotropy axis. For all curves the tip wave function on the right panel (shown in the length scale of a 2 lattice constants) was used with different relative angle $\theta$ to generate $O_N(E)$.

## TIP-CHANGING PROCEDURE ON CeCoIn$_5$ AND URu$_2$Si$_2$

The tip is offset from the measurement area and the configuration of its apex is modified by moving the tip toward the surface until we see a saturation in the tunneling current (beyond the limits of our pre-amplifier). This we call a controlled poke onto the surface. We have determined the optimal voltage applied to the scanner piezo empirically. Because of the metallic nature of the CeCoIn$_5$ and URu$_2$Si$_2$ surfaces, this controlled poke does not compromise the stability of the tunneling junction, nor does it modify the poking site beyond a few tens of angstroms. The change in the tip apex configuration is confirmed by subsequent topographic measurements showing a change in the relative intensities of the Bragg peaks.

## TIP-CHANGING PROCEDURE ON Bi$_2$Sr$_2$CaCu$_2$O$_{8+\delta}$

Contact between the tip and the Bi-2212 sample can results in an unstable tunneling junction and the possible destruction of the approached area, thus decreeing the poking procedure used in URu$_2$Si$_2$ and CeCoIn$_5$ unusable. To circumvent this problem we developed a method to change the configuration of the tip apex which preserves the stability of the tunneling junction and allows subsequent measurements on the same area. Scanning over large areas ($\approx 5000$ Å), it is possible to find small spots ($\approx 1$ Å) over which the tunneling junction is highly unstable. By scanning over them at lower bias set points ($\approx 20$ mV) while maintaining a large tunneling current ($\approx 300$ pA), the tip sample separation is decreased and the tip is more susceptible to changes in its apex configuration. Such a tip changing event is characterized by a single spike in the tunneling current. The change in the tip apex configuration is confirmed by subsequent topographic measurements.

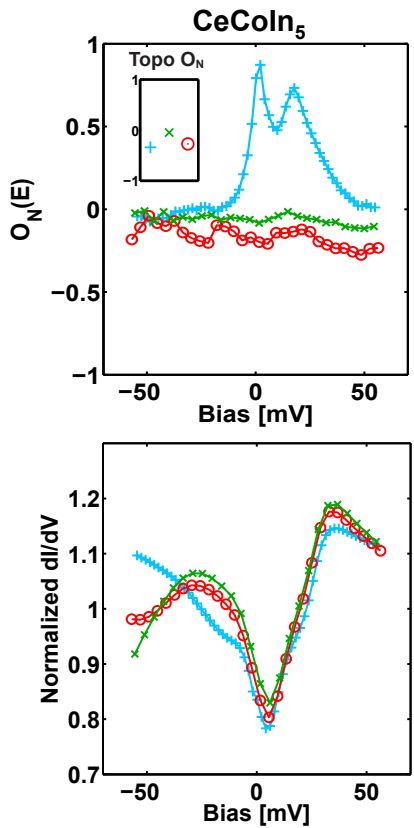

**CeCoIn₅**

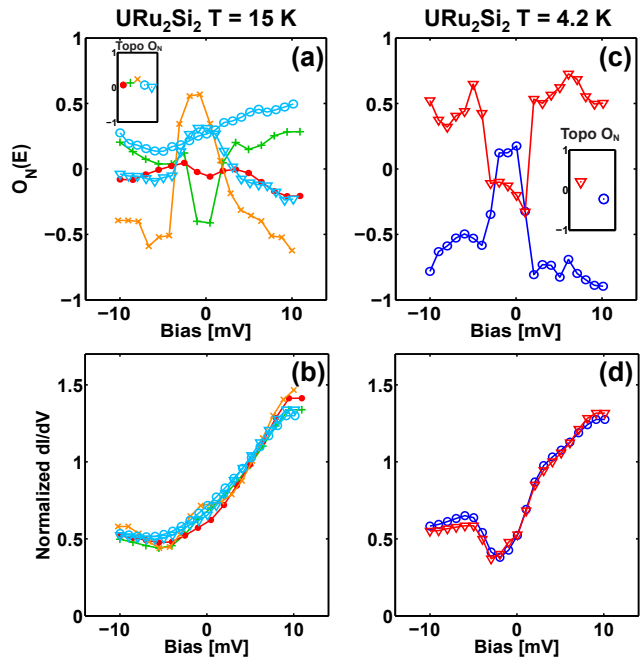

FIG. 3. (b) Average spectra from the $dI/dV$ measurements used to generate the corresponding $O_N(E)$ curves in Fig. 4(c) of the main text (reproduced in (a)). (c) $O_N(E)$ measured on URu$_2$Si$_2$ below $T_{HO}$ (4.2 K) over the same FOV with the different tips. (d) Average spectra from the $dI/dV$ measurements used to generate the corresponding $O_N(E)$ curves in (c). The insets represent the asymmetry parameter of the respective individual topographs acquired simultaneously to the conductance maps.

FIG. 2. Average spectra (lower panels) from the $dI/dV$ measurements used to generate the corresponding $O_N(E)$ curves in Fig. 4(c) of the main text (reproduced in the upper panel). The spectra were normalized to the their integral over a range of voltages ($-60 < V < 0$). The insets represent the asymmetry parameter of the respective individual topographs acquired simultaneously to the conductance maps.

## EFFECTS OF TIP-CHANGING PROCEDURE ON SPECTRA

The average spectrum for each of the measurements in Fig. 4 of the main text is displayed in Figs. 2, 3, and 4. Notice that the measurements on URu$_2$Si$_2$ displayed in Fig. 4(c) of the main text were taken just below $T_{HO}$ (15 K), where the effects of the hidden order gap on the spectrum are small [4] (see Fig. 3(b)) but where the sensitivity of $O_N(E)$ to the hidden order is strong (Fig. 3(a)). For comparison we show similar measurements taken at 4.2 K where the hidden order gap is fully developed.

Small variations between measurements are expected since the different tip-geometries should yield different tunneling matrix elements. However, the relative small variance over the different measurements indicates that tunneling spectra is mostly unchanged by the tip-changing procedures used here and that the large variance seen in the $O_N(E)$ curves displayed in Fig. 4 of the main text is likely due to different tip geometries.

## SETPOINT CONDITION FOR $O_N(E)$ MEASUREMENTS

Table I shows the setup current and bias voltage used for measuring the conductance maps from which the $O_N(E)$ curves presented in Figs. 3 and 4 of the main text are obtained.

TABLE I.

| Material | Bias (mV) | Current (pA) | Location |
|----------|-----------|--------------|----------|
| CeCoIn$_5$ | -200 | 1600 | Fig. 3 and Fig. 4(a), cyan |
| CeCoIn$_5$ | -200 | 300 | Fig. 4(a), green |
| CeCoIn$_5$ | -200 | 300 | Fig. 4(a), red |
| Bi-2212 | 400 | 200 | Fig. 4(b), all |
| URu$_2$Si$_2$ | -150 | 150 | Fig. 4(c), cyan circles |
| URu$_2$Si$_2$ | -150 | 150 | Fig. 4(c), cyan inv. triangles |
| URu$_2$Si$_2$ | -150 | 200 | Fig. 4(c), orange |
| URu$_2$Si$_2$ | -200 | 200 | Fig. 4(c), red |
| URu$_2$Si$_2$ | 100 | 200 | Fig. 4(c), green |

## DRIFT CORRECTION

All the conductance maps were corrected for drift using the analysis method outlined in Ref. [5]. This allowed the

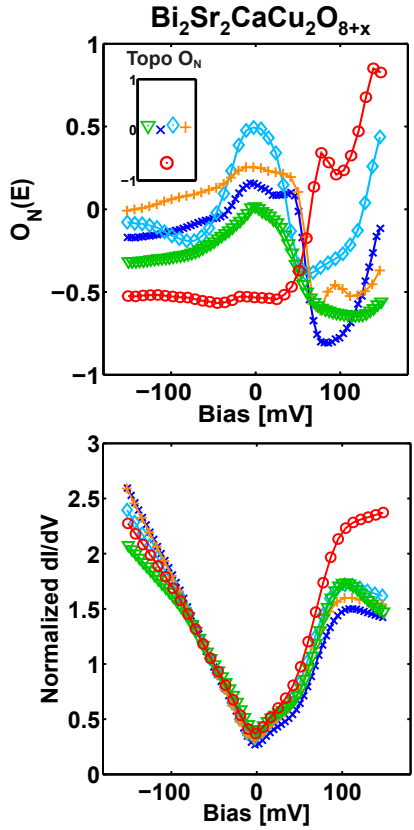

FIG. 4. Average spectra (lower panels) from the $dI/dV$ measurements used to generate the corresponding $O_N(E)$ curves in Fig. 4(b) of the main text (reproduced in the upper panel). The spectra were normalized to the their integral over a range of voltages ($-100 < V < 0$). The insets represent the asymmtery parameter of the respective individual topographs acquired simultaneously to the conductance maps.

Bragg peaks in the DFTs of the conductance maps to be represented by a single pixel. However, we find that the results presented in the current work are not dependent on the use of this analysis tool, and the same effects are present with the same magnitudes in the raw data.

---



\end{document}